\documentclass[a4paper,11pt]{article}
\usepackage{pos}
\usepackage{enumerate}
\usepackage{hyperref}

\title{Summing Feynman diagrams in the worldline formalism}

\author[a]{N. Ahmadiniaz}
\author[b]{J.P. Edwards}
\author[c]{C. Lopez-Arcos}
\author[a]{M.A. Lopez-Lopez}
\author[d]{C. Moctezuma Mata}
\author[d]{J. Nicasio} 
\author*[e]{C. Schubert}

\affiliation[a]{Helmholtz-Zentrum Dresden-Rossendorf, Bautzner Landstra\ss e 400, 01328 Dresden, Germany}
\affiliation[b]{Centre for Mathematical Sciences, University of Plymouth, Plymouth, PL4 8AA, UK}
\affiliation[c]{Escuela de Matem\'aticas, Universidad Nacional de Colombia Sede Medell\'in, Carrera 65 \# 59A-110, Medell\'in, Colombia}
\affiliation[d]{Instituto de F\'isica y Matem\'aticas
Universidad Michoacana de San Nicol\'as de Hidalgo
Edificio C-3, Apdo. Postal 2-82
C.P. 58040, Morelia, Michoac\'an, M\'exico}
\affiliation[e]{Centro Internacional de Ciencias A.C., Campus UNAM-UAEM, Cuernavaca, Morelos, M\'exico}

\emailAdd{christianschubert137@gmail.com}

\abstract{The worldline formalism shares with string theory the property that it allows one to write down master integrals that effectively combine the contributions of many Feynman diagrams. While at the one-loop level these diagrams differ only by the position of the external legs along a fixed line or loop, at multiloop they generally involve different topologies. Here we summarize various efforts that have been
made over the years to exploit this property in a computationally meaningful way. As a first example, we show how to generalize the Landau-Khalatnikov-Fradkin formula for the non-perturbative
gauge transformation of the fermion propagator in QED to the general $2n$ - point case by pure manipulations at the path-integral level. At the parameter-integral level, we show how to integrate out individual
photons in the low-energy expansion, and then sketch a recently introduced general framework for the analytical evaluation of such worldline integrals involving a reduction to quantum mechanics on the circle and the relation between inverse derivatives and Bernoulli polynomials.}

\FullConference{%
  Loops and Legs in Quantum Field Theory - LL2022,\\
  25-30 April, 2022\\
  Ettal, Germany
}

\begin{document}

\setlength{\unitlength}{1mm}

\def\nonu{\nonumber\\}

\def\ddel{{}^\bullet\! \Delta}
\def\deld{\Delta^{\hskip -.5mm \bullet}}
\def\dddel{{}^{\bullet \bullet} \! \Delta}
\def\ddeld{{}^{\bullet}\! \Delta^{\hskip -.5mm \bullet}}
\def\deldd{\Delta^{\hskip -.5mm \bullet \bullet}}
\def\epsk#1#2{\varepsilon_{#1}\cdot k_{#2}}
\def\epseps#1#2{\varepsilon_{#1}\cdot\varepsilon_{#2}}

\newcommand{\jhat}[1]{\hspace{0.3em}\widehat{\hspace{-0.4em}#1\hspace{-0.4em}}\hspace{0.4em}}
\newcommand{\bone}{1\!\!1}

%
\def\cosech{\rm cosech}
\def\sech{\rm sech}
\def\coth{\rm coth}
\def\tanh{\rm tanh}
\def\half{{1\over 2}}
\def\third{{1\over3}}
\def\fourth{{1\over4}}
\def\fifth{{1\over5}}
\def\sixth{{1\over6}}
\def\seventh{{1\over7}}
\def\eigth{{1\over8}}
\def\ninth{{1\over9}}
\def\tenth{{1\over10}}
\def\bN{\mathop{\bf N}}
\def\R{{\rm I\!R}}
\def\Eins{{\mathchoice {\rm 1\mskip-4mu l} {\rm 1\mskip-4mu l}
{\rm 1\mskip-4.5mu l} {\rm 1\mskip-5mu l}}}
\def\Z{{\mathchoice {\hbox{$\sf\textstyle Z\kern-0.4em Z$}}
{\hbox{$\sf\textstyle Z\kern-0.4em Z$}}
{\hbox{$\sf\scriptstyle Z\kern-0.3em Z$}}
{\hbox{$\sf\scriptscriptstyle Z\kern-0.2em Z$}}}}
\def\abs#1{\left| #1\right|}
\def\com#1#2{
        \left[#1, #2\right]}
\def\square{\kern1pt\vbox{\hrule height 1.2pt\hbox{\vrule width 1.2pt
   \hskip 3pt\vbox{\vskip 6pt}\hskip 3pt\vrule width 0.6pt}
   \hrule height 0.6pt}\kern1pt}
      \def\boxop{{\raise-.25ex\hbox{\square}}}
\def\contract{\makebox[1.2em][c]{
        \mbox{\rule{.6em}{.01truein}\rule{.01truein}{.6em}}}}
\def\ltap{\ \raisebox{-.4ex}{\rlap{$\sim$}} \raisebox{.4ex}{$<$}\ }
\def\gtap{\ \raisebox{-.4ex}{\rlap{$\sim$}} \raisebox{.4ex}{$>$}\ }
\def\mn{{\mu\nu}}
\def\rs{{\rho\sigma}}
\newcommand{\Det}{{\rm Det}}
\def\Tr{{\rm Tr}\,}
\def\tr{{\rm tr}\,}
\def\sumij{\sum_{i<j}}
\def\e{\,{\rm e}}
\def\non{\nonumber\\}
\def\br{{\bf r}}
\def\bp{{\bf p}}
\def\bx{{\bf x}}
\def\by{{\bf y}}
\def\brhat{{\bf \hat r}}
\def\bv{{\bf v}}
\def\ba{{\bf a}}
\def\bE{{\bf E}}
\def\bB{{\bf B}}
\def\bA{{\bf A}}
\def\pa{\partial}
\def\dA{\partial^2}
\def\ddx{{d\over dx}}
\def\ddt{{d\over dt}}
\def\der#1#2{{d #1\over d#2}}
\def\lie{\hbox{\it \$}} 
\def\partder#1#2{{\partial #1\over\partial #2}}
\def\secder#1#2#3{{\partial^2 #1\over\partial #2 \partial #3}}
\def\kinq{{1\over 4}\dot q^2}
\def\kinb{{1\over 4}\dot x^2}
%
\def\bef{\begin{frame}}
\def\ef{\end{frame}}
\def\be{\begin{equation}}
\def\ee{\end{equation}\noindent}
\def\bear{\begin{eqnarray}}
\def\ear{\end{eqnarray}\noindent}
\def\bec{\begin{equation}}
\def\eec{\end{equation}\noindent}
\def\bearc{\begin{eqnarray}}
\def\earc{\end{eqnarray}\noindent}
\def\benn{\begin{enumerate}}
\def\enn{\end{enumerate}}
\def\veject{\vfill\eject}
\def\ven{\vfill\eject\noindent}
%
\def\eq#1{{eq. (\ref{#1})}}
\def\eqs#1#2{{eqs. (\ref{#1}) -- (\ref{#2})}}
%
\def\totint{\int_{-\infty}^{\infty}}
\def\posint{\int_0^{\infty}}
\def\negint{\int_{-\infty}^0}
\def\pint{{\dps\int}{dp_i\over {(2\pi)}^d}}
%
\newcommand{\GeV}{\mbox{GeV}}
\def\FFdual{F\cdot\tilde F}
\def\bra#1{\langle #1 |}
\def\ket#1{| #1 \rangle}
\def\braket#1#2{\langle {#1} \mid {#2} \rangle}
\def\vev#1{\langle #1 \rangle}
\def\rightvac{\mid 0\rangle}
\def\leftvac{\langle 0\mid}
\def\ihbar{{i\over\hbar}}
\def\ge{\hbox{$\gamma_1$}}
\def\gz{\hbox{$\gamma_2$}}
\def\gd{\hbox{$\gamma_3$}}
\def\go{\hbox{$\gamma_1$}}
\def\gt{\hbox{\$\gamma_2$}}
\def\gth{\hbox{$\gamma_3$}} 
\def\gf{\hbox{$\gamma_5\;$}}
\def\slash#1{#1\!\!\!\raise.15ex\hbox {/}}
\newcommand{\slD}{\,\raise.15ex\hbox{$/$}\kern-.27em\hbox{$\!\!\!D$}}
\newcommand{\slpartial}{\raise.15ex\hbox{$/$}\kern-.57em\hbox{$\partial$}}

\newcommand{\PP}{\cal P}
\newcommand{\G}{{\cal G}}
\newcommand{\nc}{\newcommand}
\newcommand{\Fkala}{F_{\kappak_i\cdot k_j}}
\newcommand{\Fkanu}{F_{\kappa\nu}}
\newcommand{\Flaka}{F_{k_i\cdot k_j\kappa}}
\newcommand{\Flamu}{F_{k_i\cdot k_j\mu}}
\newcommand{\Fmunu}{F_{\mu\nu}}
\newcommand{\Fnumu}{F_{\nu\mu}}
\newcommand{\Fnuka}{F_{\nu\kappa}}
\newcommand{\Fmuka}{F_{\mu\kappa}}
\newcommand{\Fkalamu}{F_{\kappak_i\cdot k_j\mu}}
\newcommand{\Flamunu}{F_{k_i\cdot k_j\mu\nu}}
\newcommand{\Flanumu}{F_{k_i\cdot k_j\nu\mu}}
\newcommand{\Fkamula}{F_{\kappa\muk_i\cdot k_j}}
\newcommand{\Fkanumu}{F_{\kappa\nu\mu}}
\newcommand{\Fmulaka}{F_{\muk_i\cdot k_j\kappa}}
\newcommand{\Fmulanu}{F_{\muk_i\cdot k_j\nu}}
\newcommand{\Fmunuka}{F_{\mu\nu\kappa}}
\newcommand{\Fkalamunu}{F_{\kappak_i\cdot k_j\mu\nu}}
\newcommand{\Flakanumu}{F_{k_i\cdot k_j\kappa\nu\mu}}

\nc{\spa}[3]{\left\langle#1\,#3\right\rangle}
\nc{\spb}[3]{\left[#1\,#3\right]}
\nc{\ksl}{\not{\hbox{\kern-2.3pt $k$}}}
\nc{\hf}{\textstyle{1\over2}}
\nc{\pol}{\varepsilon}
\nc{\tq}{{\tilde q}}
\nc{\esl}{\not{\hbox{\kern-2.3pt $\pol$}}}
\newcommand{\cL}{\cal L}
\newcommand{\D}{\cal D}
\newcommand{\Dhalf}{{D\over 2}}
\def\eps{\epsilon}
\def\epshalf{{\epsilon\over 2}}
\def\lag{( -\partial^2 + V)}
\def\freeexp{{\rm e}^{-\int_0^Td\tau {1\over 4}\dot x^2}}
\def\kinb{{1\over 4}\dot x^2}
\def\kinf{{1\over 2}\psi\dot\psi}
\def\expk{{\rm exp}\biggl[\,\sum_{i<j=1}^4 G_{Bij}k_i\cdot k_j\biggr]}
\def\expp{{\rm exp}\biggl[\,\sum_{i<j=1}^4 G_{Bij}p_i\cdot p_j\biggr]}
\def\expshort{{\e}^{\half G_{Bij}k_i\cdot k_j}}
\def\expabb{{\e}^{(\cdot )}}
\def\epseps#1#2{\varepsilon_{#1}\cdot \varepsilon_{#2}}
\def\epsk#1#2{\varepsilon_{#1}\cdot k_{#2}}
\def\kk#1#2{k_{#1}\cdot k_{#2}}
\def\G#1#2{G_{B#1#2}}
\def\Gp#1#2{{\dot G_{B#1#2}}}
\def\GF#1#2{G_{F#1#2}}
\def\Dab{{(x_a-x_b)}}
\def\Dsq{{({(x_a-x_b)}^2)}}
\def\PITD{{(4\pi T)}^{-{D\over 2}}}
\def\4piTD{{(4\pi T)}^{-{D\over 2}}}
\def\4piT4{{(4\pi T)}^{-2}}
\def\TintmD{{\dps\int_{0}^{\infty}}{dT\over T}\,e^{-m^2T}
    {(4\pi T)}^{-{D\over 2}}}
\def\Tintm4{{\dps\int_{0}^{\infty}}{dT\over T}\,e^{-m^2T}
    {(4\pi T)}^{-2}}
\def\Tintm{{\dps\int_{0}^{\infty}}{dT\over T}\,e^{-m^2T}}
\def\Tint{{\dps\int_{0}^{\infty}}{dT\over T}}
\def\np{n_{+}}
\def\nm{n_{-}}
\def\Np{N_{+}}
\def\Nm{N_{-}}
\newcommand{\slG}{{{\dot G}\!\!\!\! \raise.15ex\hbox {/}}}
\newcommand{\Gd}{{\dot G}}
\newcommand{\Gund}{{\underline{\dot G}}}
\newcommand{\Gdd}{{\ddot G}}
\def\GBd12{{\dot G}_{B12}}
\def\Dx{\dps\int{\cal D}x}
\def\Dy{\dps\int{\cal D}y}
\def\Dpsi{\dps\int{\cal D}\psi}
\def\dint#1{\int\!\!\!\!\!\int\limits_{\!\!#1}}
\def\ddtau{{d\over d\tau}}
\def\ie{\hbox{$\textstyle{\int_1}$}}
\def\iz{\hbox{$\textstyle{\int_2}$}}
\def\id{\hbox{$\textstyle{\int_3}$}}
\def\ldop{\hbox{$\lbrace\mskip -4.5mu\mid$}}
\def\rdop{\hbox{$\mid\mskip -4.3mu\rbrace$}}
%
\newcommand{\1}{{\'\i}}
\newcommand{\no}{\noindent}
\def\non{\nonumber}
\def\dps{\displaystyle}
\def\sy{\scriptscriptstyle}
\def\sy{\scriptscriptstyle}

\maketitle

\section{QED in the worldline representation}

\subsection{Worldline representation of dressed scalar propagator}

\begin{figure}[htbp]
\begin{center}
 \includegraphics[width=0.45\textwidth]{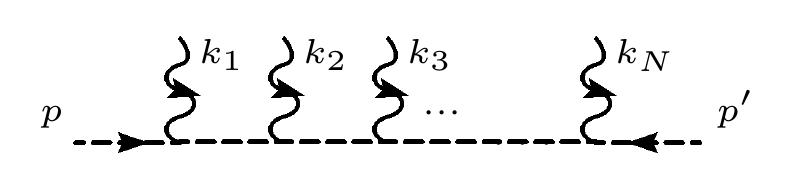}
\caption{The photon-dressed scalar propagator. }
\label{fig-propexpand}
\end{center}
\end{figure}

Let us start with Feynman's 1950 worldline path integral representation \cite{feynman1950} of the 
Green's function for the interacting Klein-Gordon operator $ -(\partial + ie A)^2 + m^2$,
\bear
D^{xx'} [A] &=& 
\langle x'| \int_0^{\infty}dT\, {\rm exp}\Bigl\lbrack - T (  -(\partial + ie A)^2+m^2)\Bigr\rbrack  | x \rangle
\nonumber\\
%
%
%
%
%
%
%
&=&
\int_0^{\infty}
dT\,
\e^{-m^2T}
\int_{x(0)=x'}^{x(T)=x}
{\cal D}x(\tau)\,
e^{-\int_0^T d\tau \bigl(\kinb +ie\dot x\cdot A(x(\tau))\bigr)}
\, .
\label{scalpropfreepi}
\ear\no
Choosing the background field as $N$ plane waves,
$
A^{\mu}(x(\tau)) = \sum_{i=1}^N \,\varepsilon_i^{\mu}\e^{ik_i\cdot x(\tau)}\, ,
$
and Fourier transforming the endpoints, we get 
a representation of the ``photon-dressed propagator'' shown in
Fig. \ref{fig-propexpand}.

Note that the worldline representation provides the complete amplitude in one piece:
differently from the Feynman diagrammatic approach, there is neither a need to sum over ``crossed'' diagrams,
nor to distinguish between the linear and the ``seagull'' vertex couplings of the scalar particle to the Maxwell field.

\subsection{Worldline representation of scalar QED effective action}

Similarly, the one-loop effective action can be presented in terms of a path integral over all closed loops in spacetime:
\bear
\Gamma [A] 
&=& \int_0^{\infty} \frac{dT}{T} \,{\rm Tr} \, {\rm exp}\Bigl\lbrack - T (  -(\partial + ie A)^2+m^2)\Bigr\rbrack  \nonumber\\
&=&
\int_0^{\infty}
\frac{dT}{T}\,
\e^{-m^2T}
\int_{x(0)=x(T)}
{\cal D}x(\tau)\,
e^{-\int_0^T d\tau \bigl(\kinb +ie\dot x\cdot A(x(\tau))\bigr)}
\, .
\label{Gammascal}
\ear
Expanding the field in $ N$ plane waves,
one gets the full scalar QED  one-loop $N$-photon amplitudes.

%

\subsection{Worldline representation of spinor QED effective action}

At the level of the $N$-photon amplitudes, the transition from scalar to spinor QED can (up to the normalization) simply be made
by inserting, under the path integral \eqref{Gammascal}, the ``Feynman spin factor''  ${\rm Spin}[x,A]$,
\bear
{\rm Spin}[x,A] = {\rm tr}_{\Gamma} {\cal P}
\exp\Biggl[{{i\over 4}e\,[\gamma^{\mu},\gamma^{\nu}]
\int_0^Td\tau F_{\mu\nu}(x(\tau))}\Biggr]\, .
\ear
However, its use implies path ordering, forcing one to fix the ordering of the photons.
For the purposes that I am going to discuss here, it will be essential to avoid this.
Following Fradkin \cite{fradkin66}, this can be done replacing the spin factor by a Grassmann path integral,
\bear
{\rm Spin}[x,A]  \to 
\int {\cal D}\psi(\tau)
\,
\exp 
\Biggl\lbrack
-\int_0^Td\tau
\Biggl(
\half \psi\cdot \dot\psi -ie \psi^{\mu}F_{\mn}\psi^{\nu}
\Biggr)
\Biggr\rbrack
\ear
where the Lorentz vectors $\psi^\mu (\tau)$ are anticommuting and antiperiodic functions of proper time:
$\psi(\tau_1)\psi(\tau_2) = - \psi(\tau_2)\psi(\tau_1), \quad\psi(T) = - \psi(0)$. 
The main point of the Grassmann approach is to replace the path-ordered exponential by an ordinary exponential.

\subsection{Worldline representation of dressed electron propagator}

For the dressed fermion propagator, too, worldline reprentations have been around for decades \cite{feynman1951,fradkin66,fragit91,18}. 
However, only during the last few years a version has been developed that seems suitable for higher-order state-of-the-art calculations \cite{130,131}.
The starting point is the second-order representation of the $x$-space Dirac propagator $S^{xx'}[A]$ in a Maxwell background:
\bear
S^{xx'}[A] &=&
\bigl[m + i\slash{D}'\bigr]
K^{xx'}[A]\,\nonumber\\
K^{xx'}[A] &=&
\Big\langle x'\Big|\Bigl[m^2- D_{\mu}D^{\mu} +{i\over 2}\, e \gamma^{\mu}\gamma^{\nu} F_{\mu\nu}\Bigr]^{-1} 
\Big| x \Big \rangle \nonumber\\
&=&
\int_0^{\infty}
dT\,
\e^{-m^2T}
\e^{-\fourth \frac{(x-x')^2}{T}}
\int_{q(0)=0}^{q(T)=0}
Dq\,
\e^{-\int_0^T d\tau\bigl(
\kinq
+ie\,\dot q\cdot A 
+ie \frac{x'-x}{T}\cdot A
\bigr)}
\nonumber\\
&& \times\,  2^{-\frac{D}{2}}
{\rm symb}^{-1}
\int_{\psi(0)+\psi(T)=0
\hspace{-30pt}}D\psi
\, \e^
{-\int_0^Td\tau\,
\bigl[\half\psi_{\mu}\dot\psi^{\mu}-ieF_{\mu\nu}(\psi+\eta)^{\mu}(\psi+\eta)^{\nu}\bigl]
}
\, .
\ear\no
Here $\eta^{\mu}$ is an external Grassmann Lorentz vector, and the ``symbol map''  {\it symb}  converts products of $\eta$s 
into fully antisymmetrised products of Dirac matrices:
\bear
{\rm symb} 
\bigl(\gamma^{\alpha_1\alpha_2\cdots\alpha_n}\bigr) \equiv 
(-i\sqrt{2})^n
\eta^{\alpha_1}\eta^{\alpha_2}\ldots\eta^{\alpha_n}
\ear
%
where $\gamma^{\alpha\beta\cdots\rho}$ denotes the totally antisymmetrised product of gamma matrices:
\bear
\gamma^{\alpha_1\alpha_2\cdots \alpha_n} \equiv \frac{1}{n!}\sum_{\pi\in S_n} 
{\rm sign}(\pi) \gamma^{\alpha_{\pi(1)}}\gamma^{\alpha_{\pi(2)}} \cdots \gamma^{\alpha_{\pi(n)}} 
\, .
\ear

\subsection{Higher order QED processes}

\begin{figure}[htbp]
\begin{center}
 \includegraphics[width=0.45\textwidth]{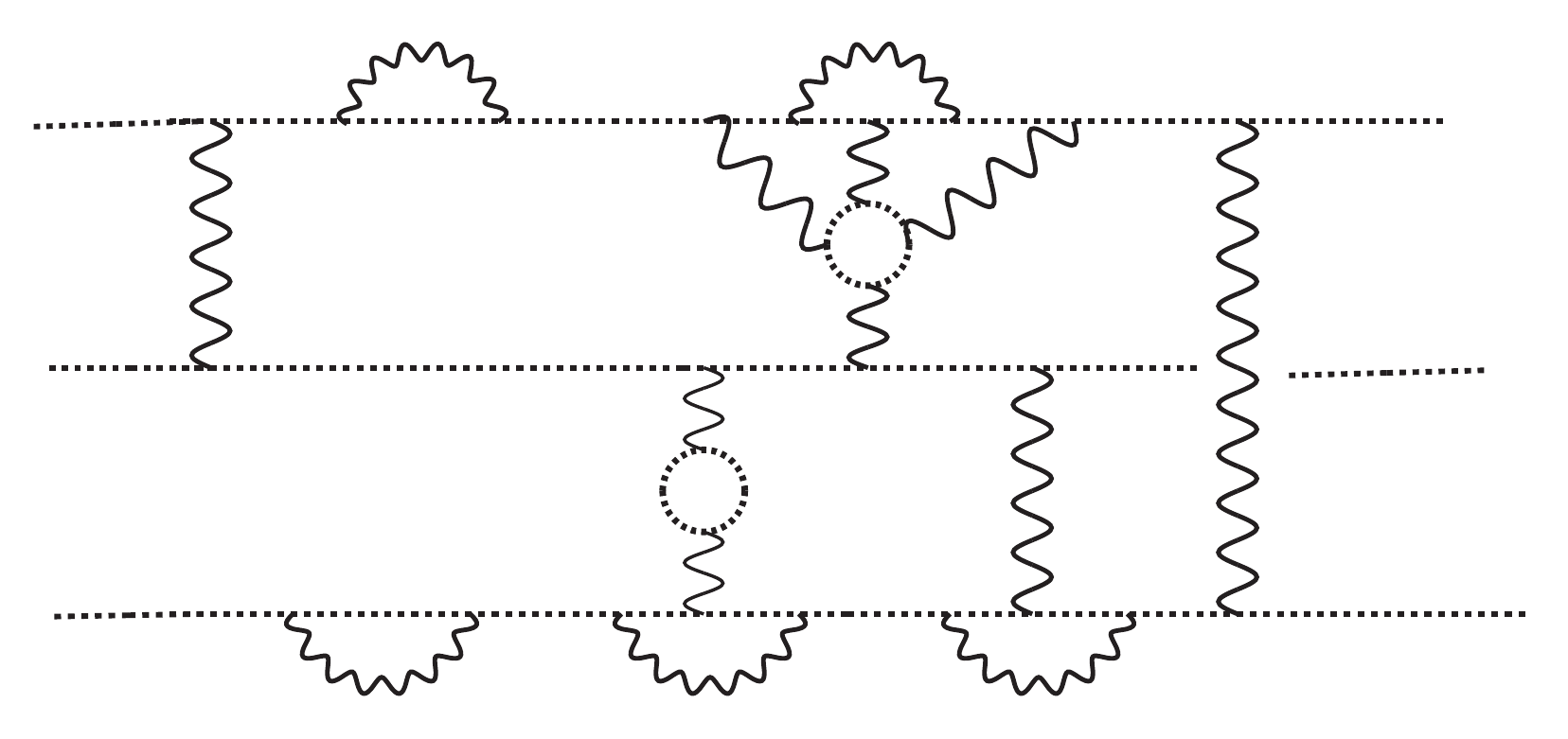}
\caption{{A typical multiloop process in QED. }}
\label{fig-QEDSmatrix}
\end{center}
\end{figure}

Since all the above formulas are valid off-shell, arbitrary QED processes  can be constructed from these building blocks by  sewing (Fig. \ref{fig-QEDSmatrix}).

Although the worldline representation is equivalent to Feynman diagrams, it is more global in the sense that it does not distinguish between diagrams differing only by the ordering of the photon legs along a line or loop.

\section{Generalization of the Landau-Khalatnikov-Fradkin transformation}

For some purposes, this property becomes useful already at the path-integral level. 
In 1956, Landau and Khalatnikov \cite{LK} and independently Fradkin \cite{fradkin} showed, that the
nonperturbative behaviour of the electron propagator $S(x; \xi)$  under a change $\xi \rightarrow \hat \xi$  of the
covariant gauge parameter $\xi$ can be written as (in exactly four dimensions)
 \begin{equation}
	S(x; \xi) = S(x; \hat{\xi})\left[\frac{x^{2}}{x^{2}_{0}}\right]^{-\frac{ \alpha}{4\pi}(\xi - \hat\xi)}
\end{equation}
where $x_0$  is an IR cutoff. Very recently, we have used the worldline formalism to extend this result to the 
general fermionic $2n$-point correlator %
\bear
\mathcal{A}(x_{1}, \ldots , x_{n}; x^{\prime}_1,\ldots, x^{\prime}_n | \xi 	) \equiv
\langle \psi(x_1) \cdots \psi(x_n)\bar\psi (x_1') \cdots \bar\psi (x_n') \rangle
\ear
as follows \cite{134}: 
\bear
&&	\mathcal{A}(x_{1}, \ldots , x_{n}; x^{\prime}_{1},\ldots, x^{\prime}_{n} | \hat \xi	) = \prod_{k, l = 1}^{n}\e^{(\hat\xi - \xi)S^{(k, l)}}\mathcal{A}(x_{1}, \ldots , x_{n}; x^{\prime}_{1},\ldots, x^{\prime}_{n} | \xi 	) 
\, ,
\\
&&	S^{(k, l)} = \frac{e^{2}}{32 \pi^{\frac{D}{2}}} \Gamma\Bigl(\frac{D}{2} \! -\! 2\Big)\Big\{ \big[\Big(x_{k} \! -\! x_{l}\Big)^{2}\big]^{2-\frac{D}{2}} \!-\! \big[\left(x_{k} - x^{\prime}_{l}\right)^{2}\big]^{2-\frac{D}{2}} 
- \big[\left(x^{\prime}_{k}\! -\! x_{l}\right)^{2}\big]^{2-\frac{D}{2}}
\nonumber\\ && \hspace{115pt}
 + \big[\left(x^{\prime}_{k} - x^{\prime}_{l}\right)^{2}\big]^{2-\frac{D}{2}}\Big\} \, .
\ear
Here the sum $k,l$ runs over all pairs of open fermion lines, and the exponential factor for fixed $k$ and $l$ implements the effect of the change of the covariant gauge parameter
for photons inserted between them in all possible ways (for photons that on one or both ends hit a closed loop there is no effect). Note also that the dimension has been left general. 

\section{String-inspired treatment of the worldline path integral}

Under the influence of string theory, in the nineties a perturbative approach to the evaluation
of worldline path integrals using worldline Green's functions was developed \cite{berkosNPB,strassler1}.
E.g. for the closed-loop case one has the basic correlators 
\bear
\langle x^{\mu}(\tau_1)x^{\nu}(\tau_2) \rangle
&=&
-G(\tau_1,\tau_2)\, \delta^{\mu\nu} , \quad
G(\tau_1,\tau_2) = \vert \tau_1 -\tau_2\vert - \frac{1}{T} \Bigl(\tau_1 -\tau_2\Bigr)^2\, , \\
\langle \psi^{\mu}(\tau_1)\psi^{\nu}(\tau_2)\rangle
&=&
G_F(\tau_1,\tau_2)\, \delta^{\mu\nu}, \quad
G_F(\tau_1,\tau_2) = {\rm sign}(\tau_1 - \tau_2)
\, .
\ear
This allows one to derive compact master formulas for the photon-dressed propagators and the one-loop $N$-photon amplitudes
in scalar and spinor QED, as well as for many other types of amplitudes (for reviews, see \cite{41,126}).

\subsection{Master formula for the N-photon amplitudes in scalar and spinor QED}

The in many ways prototypical one of these master formulas is the one for the one-loop $N$-photon amplitudes in scalar QED:
\begin{eqnarray}
\Gamma[\lbrace k_i,\varepsilon_i\rbrace]
&=&
{(-ie)}^N
{\dps\int_{0}^{\infty}}{dT\over T}
{(4\pi T)}^{-{D\over 2}}
e^{-m^2T}
\prod_{i=1}^N \int_0^T 
d\tau_i
\nonumber\\
&&\hspace{-8pt}
\times
\exp\biggl\lbrace\sum_{i,j=1}^N 
\bigl\lbrack \half G_{ij} k_i\cdot k_j
+i\dot G_{ij}k_i\cdot\varepsilon_j 
+\half\ddot G_{ij}\varepsilon_i\cdot\varepsilon_j
\bigr\rbrack\biggr\rbrace
\mid_{{\rm lin}(\varepsilon_1,\ldots,\varepsilon_N)}
\, .
\end{eqnarray}
Here $T$  is the loop proper-time and $\tau_i$  parametrizes the position of photon $i$ along the loop. 
A projection on the terms linear in each of the polarization vectors $\varepsilon_1, \ldots, \varepsilon_N$
is understood. Besides the Green's function $G_{ij}\equiv G(\tau_i,\tau_j)$ also its first and second derivatives appear,
\bear
\dot G(\tau_1,\tau_2) &=& {\rm sign}(\tau_1 - \tau_2)
- 2 {{(\tau_1 - \tau_2)}\over T}\, ,\\
\ddot G(\tau_1,\tau_2)
&=& 2 {\delta}(\tau_1 - \tau_2)
- {2\over T} \, .
\ear
\no
A similar master formula can be written down for the spinor loop using worldline supersymmetry.. However, 
in practice it is usually preferable to use an integration-by-parts that removes all the $\ddot G_{ij}$, and the 
following ``Bern-Kosower replacement rule'' \cite{berkosNPB}:
\bear
\dot G_{i_1i_2} 
\dot G_{i_2i_3} 
\cdots
\dot G_{i_ni_1}
\rightarrow 
\dot G_{i_1i_2} 
\dot G_{i_2i_3} 
\cdots
\dot G_{i_ni_1}
-
G_{Fi_1i_2}
G_{Fi_2i_3}
\cdots
G_{Fi_ni_1}\, .
\ear

\subsection{The four-photon amplitudes}

\begin{figure}[htbp]
\begin{center}
\includegraphics[scale=0.55]{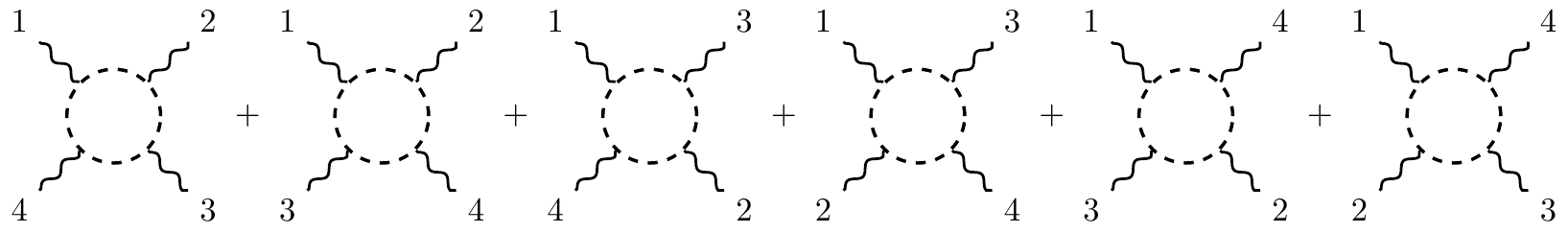}
\caption{Six Feynman diagrams for photon-photon scattering.}
\label{fig-pp}
\end{center}
\end{figure}

Let us write down the resulting representation for the familiar four-photon amplitude, usually given in terms of the
six Feynman diagrams displayed in Fig. \ref{fig-pp}.

\noindent

After a large number of integrations by parts, one finds the following tensor decomposition \cite{136} (still off-shell):
\bear
\hat \Gamma &=& \hat \Gamma^{(1)} + \hat \Gamma^{(2)}  + \hat \Gamma^{(3)}  + \hat \Gamma^{(4)}  + \hat \Gamma^{(5)} \, ,
\ear
\bear
\hat \Gamma^{(1)} &=& \hat \Gamma^{(1)}_{(1234)}T^{(1)}_{(1234)}
 + 
  \hat \Gamma^{(1)}_{(1243)}T^{(1)}_{(1243)}
   + 
  \hat \Gamma^{(1)}_{(1324)}T^{(1)}_{(1324)}
  \, ,
  \nonu
  \hat \Gamma^{(2)} &=& \hat \Gamma^{(2)}_{(12)(34)}T^{(2)}_{(12)(34)}
 + 
  \hat \Gamma^{(2)}_{(13)(24)}T^{(2)}_{(13)(24)}
   + 
  \hat \Gamma^{(2)}_{(14)(23)}T^{(2)}_{(14)(23)}
  \, ,
  \nonu
  \hat \Gamma^{(3)} &=&  \sum_{i=1,2,3} \hat \Gamma^{(3)}_{(123)i}T^{(3)r_4}_{(123)i}
  +
  \sum_{i=2,3,4} \hat \Gamma^{(3)}_{(234)i}T^{(3)r_1}_{(234)i}
  +
  \sum_{i=3,4,1} \hat \Gamma^{(3)}_{(341)i}T^{(3)r_2}_{(341)i}
  +
 \sum_{i=4,1,2} \hat \Gamma^{(3)}_{(412)i}T^{(3)r_3}_{(412)i}
 \, ,
  \nonu
  \hat \Gamma^{(4)} &=&  
\sum_{i<j}  \hat \Gamma^{(4)}_{(ij)ii}T^{(4)}_{(ij)ii} +
\sum_{i<j}  \hat \Gamma^{(4)}_{(ij)jj}T^{(4)}_{(ij)jj}
\, ,
\nonu
  \hat \Gamma^{(5)} &=&  
\sum_{i<j}  \hat \Gamma^{(5)}_{(ij)ij}T^{(5)}_{(ij)ij} +
\sum_{i<j}  \hat \Gamma^{(5)}_{(ij)ji}T^{(5)}_{(ij)ji}
\, . 
\nonumber
\ear
Remarkably, the basis of five tensors $T^{(i)}$  is identical with the one found in 1971 by Costantini, De Tollis and Pistoni \cite{cotopi} using the QED Ward identity:
\bear
T^{(1)}_{(1234)} & \equiv & Z(1234) \, ,  \\
T^{(2)}_{(12)(34)} & \equiv & Z(12)Z(34) \, , \\
T^{(3)r_4}_{(123)i} & \equiv & Z(123) \frac{r_4\cdot f_4\cdot k_i }{r_4\cdot k_4} \quad  (i=1,2,3) \, , \\
T^{(4)}_{(12)11} &\equiv & Z(12) \,\frac{k_1 \cdot f_3 \cdot f_{4} \cdot k_1}{k_3\cdot k_4} \, , \\
T^{(5)}_{(12)12} &\equiv & Z(12) \, \frac{k_1 \cdot f_3 \cdot f_{4} \cdot k_2}{k_3\cdot k_4}  .
\ear  
Here we have further introduced
\bear
f_i^{\mu\nu} &\equiv & k_i^{\mu}\varepsilon_i^{\nu} - \varepsilon_i^{\mu}k_i^{\nu}\, , \\
Z(ij)&\equiv&
\half {\rm tr}\bigl(f_if_j\bigr) = \varepsilon_i\cdot k_j\varepsilon_j\cdot k_i - \varepsilon_i\cdot\varepsilon_ik_i\cdot k_j
\, ,
\\
Z(i_1i_2\ldots i_n)&\equiv&
{\rm tr}
\Bigl(
\prod_{j=1}^n
f_{i_j}\Bigr) 
\quad (n\geq 3)
\, .
\label{defZn}
\ear\no
The corresponding parameter integrals are extremely compact:
\bear
	\hat\Gamma^{(k)}_{\cdots}
	&=& 
	\int_0^\infty \frac{dT}{T} T^{4-\frac{D}{2}}\e^{-m^2T}
	\int_0^1\prod_{i=1}^4du_i\, \hat \gamma^{(k)}_{\ldots}(\Gd_{ij})\,
	 \e^{T\sum_{i<j=1}^4G_{ij}k_i\cdot k_j}
	\label{gamma}
	\ear
where, for  spinor QED,
\bear
\hat \gamma^{(1)}_{(1234)} &=& \Gd_{12}\Gd_{23}\Gd_{34}\Gd_{41} - G_{F12}G_{F23}G_{F34}G_{F41}\, , \\
\hat \gamma^{(2)}_{(12)(34)} &=& \bigl(\Gd_{12}\Gd_{21} - G_{F12}G_{F21}\bigr)  \bigl(\Gd_{34}\Gd_{43} - G_{F34}G_{F43}\bigr)\, , \\
\hat \gamma^{(3)}_{(123)i} &=& \bigl(\Gd_{12}\Gd_{23}\Gd_{31} - G_{F12}G_{F23}G_{F31}\bigr) \Gd_{4i}\, ,\\
\hat \gamma^{(4)}_{(12)11} &=&  \bigl(\Gd_{12}\Gd_{21} - G_{F12}G_{F21}\bigr)  \Gd_{13}\Gd_{41}\, , \\
\hat \gamma^{(5)}_{(12)12} &=&  \bigl(\Gd_{12}\Gd_{21} - G_{F12}G_{F21}\bigr)  \Gd_{13}\Gd_{42} 
\label{hatgamma}
\ear
(plus permutations thereof), and the coefficient functions for  scalar QED are obtained from these simply by deleting all the $G_{Fij}$. 

\subsection{Master formula for the N-photon dressed scalar propagator}

A master formula for the scalar propagator dressed with $N$ photons was obtained by Daikouji et al. in 1996 \cite{dashsu}:
\bear 
D^{pp'}(k_1,\varepsilon_1;\cdots; k_N,\varepsilon_N)&=&(-ie)^N
\int_0^\infty dT\,{\rm e}^{-m^2T}\nonumber\\
&&\hspace{-1.5cm}\times \prod_{i=1}^N\int_0^Td\tau_i\,
{\rm e}^{-Tb^2+\sum_{i,j=1}^N[\Delta_{ij}k_i\cdot k_j-2i\ddel_{ij}\varepsilon_i\cdot k_j-\ddeld_{ij}\varepsilon_i\cdot \varepsilon_j]}\Big\vert_{\varepsilon_1\varepsilon_2\cdots\varepsilon_N}\, .\nonumber\\
\ear
Here we have introduced the vector $b \equiv p'+\frac{1}{T}\sum_{i=1}^N(k_i\tau_i-i\varepsilon_i)$
and a different worldline Green's function $\Delta(\tau,\tau')$,
\bear
\langle q^\mu (\tau) q^{\nu}(\tau') \rangle &=& - 2 \Delta(\tau,\tau') \delta^\mn \, , \quad
\Delta(\tau,\tau')=\frac{\vert\tau-\tau'\vert}{2}-\frac{\tau+\tau'}{2}+\frac{\tau\tau'}{T} 
\,. 
\ear 

\subsection{Master formula for the photon-dressed electron propagator}

To the contrary, a master formula for the photon-dressed electron propagator
was obtained only very recently \cite{130,131}:
\bear
K^{pp'}_N(k_1,\varepsilon_1;\ldots ; k_N,\varepsilon_N) &=&
(-ie)^N 
{\rm symb}^{-1}
\int_0^{\infty}
dT
\e^{-m^2T}
 \int_0^Td\tau_1   \cdots  \int d\theta_N
 \nonumber\\
&& \hspace{-100pt} \times 
\e^{ - \sqrt{2}\eta \cdot \sum_{i=1}^N (\varepsilon_i+ i \theta k_i)
+\sum_{i,j=0}^{N+1} \bigl[\jhat g_{ij}k_i\cdot k_j +2iD_i\jhat g_{ij}\varepsilon_i\cdot k_j  + D_iD_j \jhat g_{ij} \varepsilon_i\cdot\varepsilon_j\bigr] }
\Big\vert_{\varepsilon_1\cdots \varepsilon_N}\, .
\ear\no
This now involves Grassmann variables $\theta_1, \ldots,\theta_N$, the super derivative $D =  {\partial\over{\partial\theta}} - \theta{\partial\over{\partial\tau}}$
and the super worldline Green's function $\jhat g(\tau,\theta;\tau',\theta') =\half (\vert \tau -\tau' \vert  + \theta\theta' {\rm sign}(\tau - \tau') )$. 
The main advantages of this approach are that (i) the use of the symbol map leads to an early projection on the Clifford basis, effectively avoiding the
appearance of long Dirac traces (ii) the spin-averaging can be done without fixing the  number or helicity assignments of the photons. 
As a check, in \cite{131} the formalism was used to recalculate the (polarized and unpolarized) Compton scattering amplitude, and complete
agreement was found with \cite{dendit-compton}. 

\subsection{On to multiloop}

\begin{figure}[htbp]
\begin{center}
\includegraphics[scale=0.35]{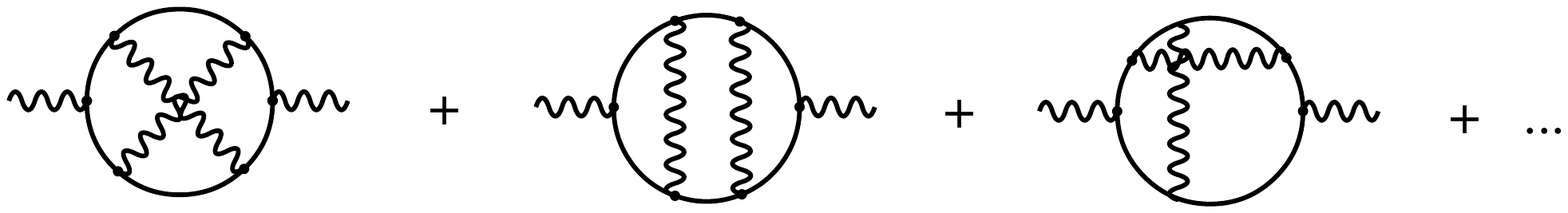}
\vspace{-120pt}
\caption{The quenched three-loop photon propagator.}
\label{fig-3loopbetadiag}
\end{center}
\end{figure}

Dealing with the amplitude as a whole becomes important when one uses the one-loop amplitudes
to construct higher-loop amplitudes by sewing. For example, from the one-loop six-photon amplitude 
we can construct the three-loop quenched propagator (Fig. \ref{fig-3loopbetadiag}) etc.

%
%

\vspace{-120pt}

\noindent

This type of sums of diagrams is known to suffer from extensive cancellations.

\section{The fundamental problem of worldline integration}

Returning to the one-loop level, considering that $G_{Fij}$s can always be eliminated by
\bear
G_{Fij}G_{Fjk}G_{Fki} = - (\dot G_{ij} + \dot G_{jk} + \dot G_{ki})  
\ear
 the most general integral that one will ever have to compute in the worldline approach to QED
(or any abelian theory)
is of the form
\bear
\int_0^1du_1du_2 \cdots du_N \, {\rm Pol} (\dot G_{ij}) \,\e^{\sum_{i<j=1}^N  G_{ij} k_i\cdot k_j}
\ear
with arbitrary $ N$ and polynomial ${\rm Pol}(\dot G_{ij})$. Here we have rescaled $\tau_i = T u_i$, 
so that now
\bear
G_{ij} = |u_i-u_j| - (u_i-u_j)^2, \quad
\dot G_{ij} = {\rm sign}(u_i-u_j) - 2(u_i-u_j) \, .
\ear
Ideally, we would like to compute this integral without decomposing the integrand into ordered sectors. 
This leads to a very non-standard integration problem. 
An easily solvable special case are ``cycle integrals'':
\bear
b_n &\equiv& \int_0^1 du_1du_2\ldots du_n\,
\dot G_{12}\dot G_{23}\cdots\dot G_{n1} =
\qquad\left\{ \begin{array}{r@{\quad\quad}l}
-2^n{{B}_n\over n!}  & \qquad n{\rm \quad even}\\
0 & \qquad n{\rm \quad odd}\\
\end{array} \right.
\ear
where ${B}_n$ denotes the $n$th Bernoulli number, and ``super cycle integrals'', 
\bear
b_n-f_n &\equiv& \int_0^1 du_1du_2\ldots du_n\,
\Bigl(\dot G_{B12}\dot G_{B23}\cdots\dot G_{Bn1} 
-
G_{F12}G_{F23}\cdots G_{Fn1}\Bigr)
=
(2-2^n)\,b_n 
\, .
\nonumber\\
\ear
In the worldline formalism, those are all that is needed to calculate the one-loop $N$-photon amplitudes in the low-energy approximation \cite{51}. 

\begin{figure}[htbp]
\begin{center}
\includegraphics[scale=0.35]{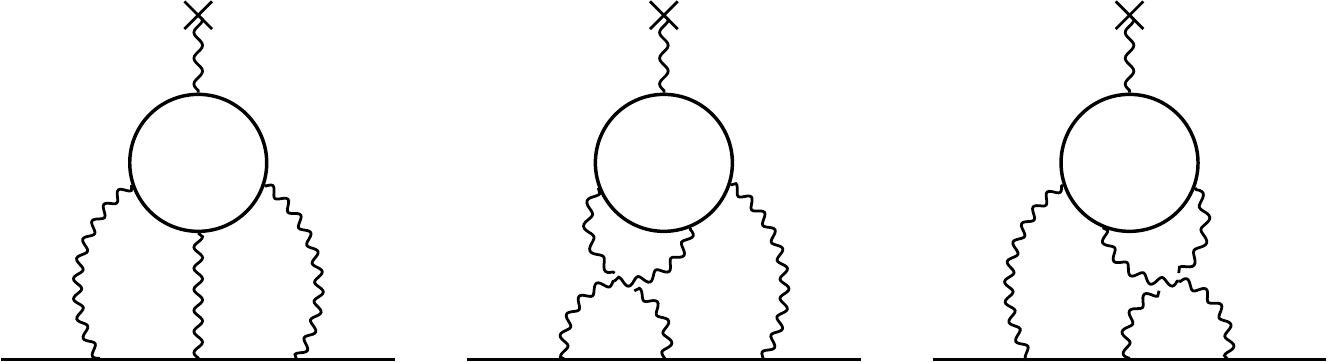}
\vspace{20pt}
\caption{LBL contributions to the three-loop $g-2$.}
\label{fig-3loop_g-2}
\end{center}
\end{figure}

\begin{figure}[htbp]
\begin{center}
\includegraphics[scale=0.35]{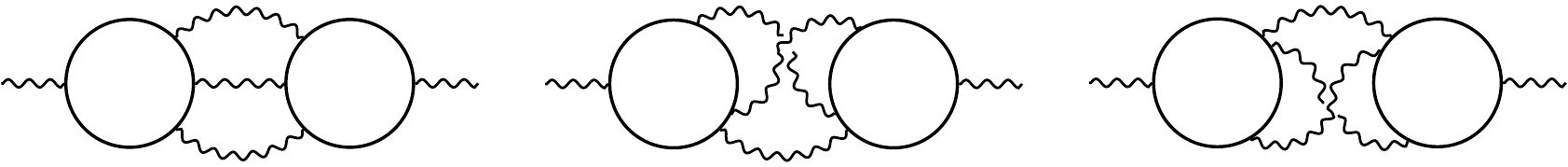}
\vspace{20pt}
\caption{LBL contributions to the four-loop $\beta$ function. The external photons are at low-energy.}
\label{fig-4loop}
\end{center}
\end{figure}

General polynomial integrals can be done recursively by the application of the following formula \cite{135}
\bear
\int_0^1 du\,
\dot G(u,u_1)^{k_1}
\dot G(u,u_2)^{k_2}
\cdots
\dot G(u,u_n)^{k_n}
&=&
{1\over 2n}
\sum_{i=1}^n\,
\prod_{j\ne i}
\sum_{l_j=0}^{k_j}
{k_j\choose l_j}
\dot G_{ij}^{k_j-l_j}
\sum_{l_i=0}^{k_i}{k_i\choose l_i}
\non\\&&\hspace{-220pt}\times
{(-1)^{\sum_{j=1}^n l_j}
\over (1+\sum_{j=1}^n l_j)n^{\sum_{j=1}^n l_j}}
\biggr\lbrace
\Bigl( \sum_{j\ne i}\dot G_{ij} +1 \Bigr)
^{1+\sum_{j=1}^n l_j}
- (-1)^{k_i-l_i}
\Bigl(
\sum_{j\ne i}\dot G_{ij} -1
\Bigr)^{1+\sum_{j=1}^n l_j }
\biggr\rbrace
\, .
\nonumber\\
\ear
This can be used, for example, to integrate out a low-energy photon in a multiphoton
amplitude {\it without fixing the order of the remaining photons}. In a forthcoming publication
\cite{LBL2} this is applied to the four-photon amplitude, creating a building block for higher-loop calculations that will
make it possible, for example, to unify the calculation of the various 3-loop g-2 contributions shown in Fig. \ref{fig-3loop_g-2}
and the calculation of the 4-loop $\beta$-function contributions shown in Fig. \ref{fig-4loop}. 

%


\vspace{-10pt}

\section{Inverse derivative expansion}

Much more difficult is the inclusion of the universal factors $\,\e^{G_{ij} k_i\cdot k_j}$. 
The following strategy was proposed in \cite{135}: the worldline path integral is performed in 
the Hilbert space $ H_P'$ of periodic functions orthogonal to the constant functions (because the zero mode must be fixed). 
In this space the ordinary $ n$th derivative $\partial_P$ is invertible, and the integral
kernel of the inverse is given essentially by the $ n$th Bernoulli polynomial $ B_n(x)$ \cite{5,15}:
\bear
<u_1\mid {\partial}^{-n}_P \mid u_{n+1}>
&=& -{1\over n!}B_n(\vert u_1-u_{n+1}\vert)
{\rm sign}^n(u_1-u_{n+1})
\quad (n\geq 1)
\, ,\\
\bra{u_i} \partial^{0}\ket{u_j} &=& \delta(u_i-u_j) -1
\, .
\ear
Thus worldline integration naturally relates to the theory of  Bernoulli numbers and polynomials.
Further, in \cite{135} it was shown how to expand the universal exponential factor in terms of the matrix elements
of the inverse derivatives:
\begin{eqnarray}
\e^{G_{ij} k_i\cdot k_j}
=1+2 \sum_{n=1}^\infty (k_i\cdot k_j)^{n-1/2}H_{2n-1} \left( \frac{\sqrt{k_i\cdot k_j}}{2}\right)
(\bra{u_i} \partial^{-2n}\, \ket{u_j} - \bra{u_i} \partial^{-2n}\, \ket{u_i} )
\end{eqnarray}
where the $H_n(x)$ are Hermite polynomials.
Let us show how to use this expansion for the simplest case of the $N$-point function in $\phi^3$ theory,
where the master formula simply becomes ($k_{ij} \equiv k_i\cdot k_j$)
\begin{eqnarray}
\hat I_N(k_1,\ldots,k_N)
&=&\int_0^\infty \frac{dT}{T}T^{N-D/2} e^{-m^2 T} \int_0^1 du_1 . . . du_N \, \mathrm{exp} \left[ T\sum_{i<j=1}^N G_{ij} k_{ij}\right]
\, .
\end{eqnarray}
In the three-point case one can use this to write ($\hat B_n \equiv \frac{B_n}{n!}$)
\begin{eqnarray}
e^{k_{12} G_{12}+k_{13} G_{13}+k_{23} G_{23}}&=&
 \left\{1+2 \sum_{i=1}^\infty k_{12}^{i-\frac{1}{2}}H_{2i-1} \left( \frac{\sqrt{k_{12}}}{2}\right)
  \left[ \bra{u_1} \partial^{-2i} \ket{u_2} + \hat{B}_{2i}\right] \right\} 
 \nonumber\\ &&  \times
 \left\{1+2 \sum_{j=1}^\infty k_{13}^{j-\frac{1}{2}}H_{2j-1} \left( \frac{\sqrt{k_{13}}}{2}\right) \left[ \bra{u_1} \partial^{-2j}\ket{u_3} + \hat{B}_{2j} \right]\right\}
\nonumber
\\
&& \times
\left\{ 1+2 \sum_{k=1}^\infty k_{23}^{k-\frac{1}{2}}H_{2k-1} \left( \frac{\sqrt{k_{23}}}{2}\right) \left[ \bra{u_2} \partial^{-2k} \ket{u_3} + \hat{B}_{2k}\right] \right\}
\, . \nonumber\\
\end{eqnarray}
Since $\int_0^1 du_{i,j}  \bra{u_i} \partial^{-2i} \ket{u_j} =0$, the three  
$\bra{u_i} \partial^{-2n} \ket{u_j}$ can produce a non-vanishing integral only together, and then by $\int_0^1du \ket{u} \bra{u} = \Eins$, 
\begin{eqnarray}
\int_{123}  \bra{u_1} \partial^{-2i}\ket{u_2} \bra{u_2} \partial^{-2k}\ket{u_3}  \bra{u_3} \partial^{-2j}\ket{u_1} &=&
{\rm Tr}( \partial^{-2(i+j+k)}) =
 -\hat{ B}_{2(i+j+k)}
 \, .
\end{eqnarray}
In this way we get a closed form-expression for the $ N=3$ momentum expansion coefficients, 
\begin{eqnarray}
I_3(a,b,c)\equiv
\int_{123} G_{12}^a G_{13}^b G_{23}^c &=& a! b! c! \sum_{i=\lfloor 1+a/2 \rfloor}^a \sum_{j=\lfloor 1+b/2 \rfloor}^b \sum_{k=\lfloor 1+c/2 \rfloor}^c h_i^a h_j^b h_k^c
\nonumber\\ && \times
\left(  \hat{B}_{2i}\hat{B}_{2j}\hat{B}_{2k}- \hat{B}_{2(i+j+k)} \right)
\, .
\end{eqnarray}
Here we have assumed that $ a,b,c$ are all different from zero, and the coefficients $ h_i^a$ are 
(from the explicit formula for the Hermite polynomials)
\bear
h_i^a = (-1)^{a+1} \frac{2(2i-1)!}{(2i-a-1)!(2a-2i+1)!}
\, .
\ear
At the four-point level, we encounter more general integrals involving  the ``cubic worldline vertex'' 
\bear
V_{3}^{ijk} &\equiv& \int_0^1 du \bra{u} \partial^{-i}\ket{u_1} \bra{u} \partial^{-j}\ket{u_2} \bra{u} \partial^{-k}\ket{u_3} 
\ear
but they can be reduced to chain integrals by a systematic IBP procedure. This remains true at higher points.

\section{Summary and Outlook}

\benn

\item
In the worldline formalism, we can integrate out photons in the low-energy limit, or to any finite
order in the external momentum, without fixing the ordering of the remaining legs.

\item
At full momentum, we can use the inverse derivative expansion, and try a resummation. For this purpose,
eventually we will need formulas relating the Bernoulli numbers to hypergeometric functions. 

\item
This provides also a powerful new approach to the calculation of the $\phi^3$ and QED heat-kernel expansions.

\enn

\end{document}